\documentclass{webofc}

\usepackage[varg]{txfonts}   
\newcommand {\snn}	{\sqrt{s_{_{\rm NN}}}}
\newcommand{ \Dg}{\Delta\gamma}
\newcommand {\Bvec}	{\vec{B}}
\newcommand {\pA}	{$p$+Au}
\newcommand {\dA}	{$d$+Au}

\newcommand {\RuRu}	{Ru+Ru}
\newcommand {\ZrZr}	{Zr+Zr}
\newcommand {\AuAu}	{Au+Au}
\newcommand {\pPb}	{$p$+Pb}
\newcommand {\PbPb}	{Pb+Pb}

\begin{document}
\title{Scaling properties of background- and chiral-magnetically-driven charge separation: evidence for detection of the chiral magnetic effect in heavy ion collisions
}
%
%

\author{\firstname{Roy} \lastname{Lacey}\inst{1}\fnsep\thanks{\email{Roy.Lacey@Stonybrook.edu}} 
}

\institute{Depts. of Chemistry \& Physics, Stony Brook University, Stony Brook, New York 11794, USA
          }
\abstract{
  The scaling properties of the $R_{\Psi_2}(\Delta S)$ correlator and the $\Delta\gamma$ correlator are used to investigate a possible chiral-magnetically-driven (CME) charge separation in $p$+Au, $d$+Au, Ru+Ru, Zr+Zr, and Au+Au collisions at $\sqrt s_{\mathrm{NN}}=200$~GeV, and in $p$+Pb ($\sqrt s_{\mathrm{NN}}=5.02$~TeV) and Pb+Pb collisions at $\sqrt s_{\mathrm{NN}}=5.02$ and $2.76$~TeV. The results for $p$+Au, $d$+Au, $p$+Pb, and Pb+Pb collisions, show the $1/{\rm N_{ch}}$ scaling for background-driven charge separation. However,  the results for Au+Au, Ru+Ru, and Zr+Zr collisions show scaling violations which indicate a CME contribution in the presence of a large background. In mid-central collisions, the CME accounts for approximately 27\% of the signal + background in Au+Au and roughly a factor of two smaller for Ru+Ru and Zr+Zr, which show similar magnitudes.
}
%
%
%
\maketitle
%
%
Metastable domains of gluon fields with non-trivial topological configurations can form in the magnetized chiral relativistic quark-gluon plasma (QGP) \cite{Kharzeev:2004ey} produced in collisions at RHIC and the LHC. The colliding ions generate the magnetic field ($\Bvec$) at early times~\cite{Asakawa:2010bu}. The interaction of chiral quarks with the gluon fields can drive a chiral imbalance resulting in an electric current  
%
$
\vec{J}_V = \frac{N_{c}e\vec{B}}{2\pi^2}\mu_A, 
$
along the $\Bvec$-field, i.e., roughly perpendicular to the reaction plane; $N_c$ is the color factor, and $\mu_A$ is the axial chemical potential that quantifies the imbalance between right- and left-handed quarks. The resulting final-state charge separation, termed the chiral magnetic effect (CME)~\cite{Kharzeev:2004ey}, is of great experimental and theoretical interest. However, its experimental characterization has been hampered by significant background.

The charge separation can be quantified via the $P$-odd sine term ${a_{1}}$, in the Fourier decomposition of the charged-particle azimuthal 
distribution~\cite{Voloshin:2004vk}:
\begin{eqnarray}\label{eq:a1}
{\frac{dN_{\rm ch}}{d\phi} \propto 1 + 2\sum_{n} (v_{n} \cos(n \Delta\phi) + a_n \sin(n \Delta\phi)  + ...)}\
\end{eqnarray}
where $\mathrm{\Delta\phi = \phi -\Psi_{RP}}$ gives the particle azimuthal angle with respect to the reaction plane (${\rm RP}$) angle, and ${v_{n}}$ and ${a_{n}}$ denote the coefficients of the $P$-even and $P$-odd Fourier terms, respectively. A direct measurement of $a_1$, is not possible due to the strict global $\cal{P}$ and $\cal{CP}$ symmetry of QCD. However, their fluctuation and/or variance $\tilde{a}_1= \left<a_1^2 \right>^{1/2}$ can be measured with charge-sensitive  correlators such as the $\gamma$-correlator~\cite{Voloshin:2004vk}  and the ${R_{\Psi_2}(\Delta S)}$ 
correlator~\cite{Magdy:2017yje}.

The $\gamma$-correlator measures charge separation as:
$
%
\gamma_{\alpha\beta} = \left\langle \cos\big(\phi_\alpha +
\phi_\beta -2 \Psi_{2}\big) \right\rangle, \nonumber \quad
\Delta\gamma = \gamma_{\rm OS} - \gamma_{\rm SS},
\label{eq:2}
$
where $\Psi_{2}$ is the azimuthal angle of the $2^{\rm nd}$-order event plane which fluctuates about  the ${\rm RP}$, $\phi$ denote the particle azimuthal emission angles, $\alpha,\beta$ denote the electric charge $(+)$ or $(-)$ and SS and OS represent same-sign ($++,\,--$) and opposite-sign ($+\,-$) charges. Measurements of the quotient $\Delta\gamma/v_2$ with the $2^{\rm nd}$-order anisotropy coefficient $v_2$, are usually employed to aid quantification of the background-driven charge separation. 

The $R_{\Psi_2}(\Delta S)$ correlator measures charge separation relative to $\Psi_2$ via the ratio:
%
$
R_{\Psi_2}(\Delta S) = C_{\Psi_2}(\Delta S)/C_{\Psi_2}^{\perp}(\Delta S), 
\label{eq:4}
$
where $C_{\Psi_2}(\Delta S)$ and $C_{\Psi_2}^{\perp}(\Delta S)$ are correlation functions that quantify charge separation $\Delta S$, approximately parallel and perpendicular (respectively) to the $\vec{B}$-field. The charge-shuffling procedure used to construct the correlation functions ensures identical properties for their numerator and denominator, except for the charge-dependent correlations, which are of interest~\cite{Magdy:2017yje}. $C_{\Psi_2}(\Delta S)$ measures both CME- and background-driven charge separation while $C_{\Psi_2}^{\perp}(\Delta S)$ measures only the background.  After correcting the $R_{\Psi_2}(\Delta S)$ distributions for the effects of particle-number fluctuations and the event-plane resolution, their inverse variance ${\sigma^{-2}_{R_{\Psi_2}}}$ are used to quantify the charge separation~\cite{Magdy:2017yje}.

In this work, we use model simulations to chart the scaling properties of ${\sigma^{-2}_{R_{\Psi_2}}}$ and $\Dg/v_2$  for the background and signal + background, respectively, in A+A collisions. We then leverage these scaling properties to  identify and characterize a possible CME-driven charge separation using previously published data for \pA, \dA, \RuRu, \ZrZr\ and \AuAu\ collisions  at RHIC~\cite{STAR:2009tro,STAR:2009wot,STAR:2013ksd,STAR:2014uiw,STAR:2019xzd,STAR:2021mii}, and \pPb\ and \PbPb\ collisions at the LHC~\cite{ALICE:2012nhw,CMS:2016wfo,CMS:2017lrw,ALICE:2017sss}.

\begin{figure}[t]
\centering
\includegraphics[width=0.78\linewidth, angle=-00]{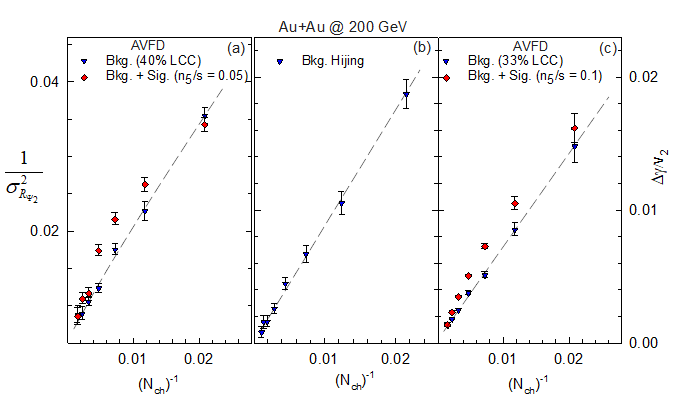}
\vskip -0.10in
\caption{ ${\sigma^{-2}_{R_{\Psi_2}}}$ vs. $1/{\rm N_{ch}}$ (a)  and  $\Dg/v_2$ vs. $1/{\rm N_{ch}}$ (b) and (c) for simulated Au+Au collisions at $\sqrt s_{\mathrm{NN}}=200$~GeV. The results for ${\sigma^{-2}_{R_{\Psi_2}}}$ and $\Dg/v_2$ from the AVFD model are shown for background and  for signal + background as indicated. The Hijing model results are only shown for the background. The dashed lines represent linear fits to the background values.
} 
\label{fig1} 
\end{figure}

Figure~\ref{fig1} shows the results for ${\sigma^{-2}_{R_{\Psi_2}}}$ and $\Dg/v_2$ obtained with the AVFD and Hijing models for Au+Au collisions. Note that these models emphasize different sources for the charge-dependent non-flow background. The AVFD model also allows input values for the initial axial charge density $n_5/s$ and the degree of local charge conservation (LCC) that regulate the magnitude of the CME- and background-driven charge separation.
%
%
The solid triangles in Fig.~\ref{fig1} show that the background [for both models]  scales as  $1/{\rm N_{ch}}$ -- the expected trend for the charge-dependent non-flow correlations. By contrast, the signal (Sig.) + background values (solid diamonds) indicate positive deviations from the background scaling \cite{Lacey:2022baf,Lacey:2022plw}. This dependence can be represented as;
%
$
  {\Delta\gamma}/{v_2}  = a + {b}/{(\rm N_{ch})^{1- c}},
   \label{eq:5}
	$
and 
%
$
   {\sigma^{-2}_{R_{\Psi_2}}}  = a' + {b'}/{(\rm N_{ch})^{1- c'}},
   \label{eq:5}
	$
%
for the small values of $n_5/s$ indicated in Fig.~\ref{fig1}.
Here, $a$, $b$ and $c$ are parameters; $c$ characterizes the degree of the scaling violation and the magnitude of the CME; for $c \sim 0$ the $1/{\rm N_{ch}}$ scaling for the background is retrieved,  as demonstrated for ${\sigma^{-2}_{R_{\Psi_2}}}$ and $\Dg/v_2$ with the AVFD model in Fig.~\ref{fig1}. 

The scaling violation for ${\sigma^{-2}_{R_{\Psi_2}}}$ and $\Dg/v_2$  (Figs.~\ref{fig1} (a) and (c)) gives a direct signature of the CME-driven contributions to the charge separation. It can be quantified via $c > 0$ and the fractions:
%
	%
$
	f^{\Dg}_{\rm CME} ={[{\Dg/v_2}(Sig.+Bkg.)-{\Dg/v_2}(Bkg.)]}/
{[{\Dg/v_2}(Sig.+Bkg.)]}, \,  f^{R}_{\rm CME} ={[{{\sigma^{-2}_{R_{\Psi_2}}}}(Sig.+Bkg.)-{{\sigma^{-2}_{R_{\Psi_2}}}}(Bkg.)]}/
{[{{\sigma^{-2}_{R_{\Psi_2}}}}(Sig.+Bkg.)]}. 
$
The scaling patterns in Fig. \ref{fig1} suggest that the observation of $1/{\rm N_{ch}}$ scaling for the experimental ${\sigma^{-2}_{R_{\Psi_2}}}$ and $\Dg/v_2$ measurements would strongly indicate background-driven charge separation with little room for a CME contribution. However, observing a violation of this $1/{\rm N_{ch}}$ scaling would indicate the CME-driven contribution. Figs. \ref{fig1} (a) and (c) also indicate comparable background and signal + background ${\sigma^{-2}_{R_{\Psi_2}}}$ and $\Dg/v_2$ values in central and peripheral collisions, suggesting that the background dominates over that of the CME-driven contributions in these collisions. This trend for the background is consistent with the reduction of $\Bvec$ in central collisions and the enhanced de-correlation between the event plane and the $\Bvec$-field in peripheral collisions. 

Since the background dominates in central and peripheral collisions, the ${\sigma^{-2}_{R_{\Psi_2}}}$ and $\Dg/v_2$ measurements for these collisions can be leveraged with $1/{\rm N_{ch}}$ scaling to obtain a quantitative estimate of the background over the entire centrality span (cf. Fig. \ref{fig1}). Here, an important proviso is to experimentally establish that the background in $p(d)$+A and A+A collisions scale as $1/{\rm N_{ch}}$  over the full centrality span. 
			


%
%
\begin{figure*}[h]
\centering
\includegraphics[width=0.83\linewidth, angle=-00]{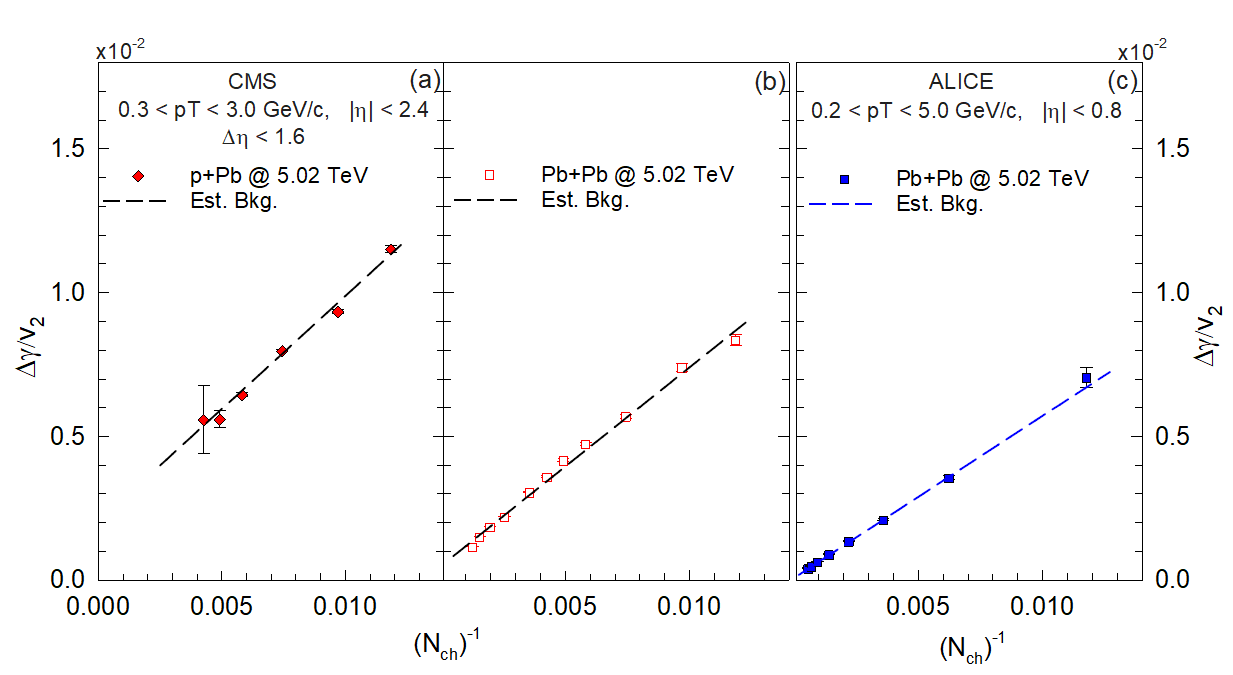}
\vskip -0.10in
\caption{ $\Dg/v_2$ vs. $1/{\rm N_{ch}}$ for \pPb\ (a) and \PbPb\ [(c) and (d)] collisions at $\sqrt s_{\mathrm{NN}}=5.02$~TeV. The dashed lines indicate an estimate of the background. The data are taken from Refs.~\cite{CMS:2016wfo,CMS:2017lrw,ALICE:2020siw}.
} 
\label{fig2} 
\end{figure*} 
The $v_2$ and $\Dg$ values reported for \pA, \dA, \RuRu, \ZrZr\ and \AuAu\ collisions  at RHIC~\cite{STAR:2009tro,STAR:2009wot,STAR:2013ksd,STAR:2014uiw,STAR:2019xzd,STAR:2021mii}, and \pPb\ and \PbPb\ collisions at the LHC~\cite{ALICE:2017sss,CMS:2016wfo,CMS:2017lrw,ALICE:2020siw} were used to investigate the scaling properties of $\Dg/v_2$. Fig.~\ref{fig2} shows the results for \pPb\ and \PbPb\ collisions at $\sqrt s_{\mathrm{NN}}=5.02$~TeV. They indicate that $\Dg/v_2$ essentially scales as $1/{\rm N_{ch}}$ ($c \approx 0$), suggesting negligible CME contributions in these collisions. They also confirm that the combined sources of background (LCC, resonances, back-to-back jets, ...), which should be substantial, especially for \pPb, scale as $1/{\rm N_{ch}}$. Note as well that the CME contribution is negligible in $p(d)$+A collisions because of significant reductions in $\Bvec$, and the sizable de-correlation between the event plane and the $\Bvec$-field~\cite{CMS:2016wfo}. Thus, the scaling patterns of $\Dg/v_2$ for these systems’ sizable backgrounds give a direct experimental constraint on the validity of  $1/{\rm N_{ch}}$ scaling of the background.

%
%
\begin{figure*}[tb]
\centering
\includegraphics[width=0.75\linewidth, angle=-00]{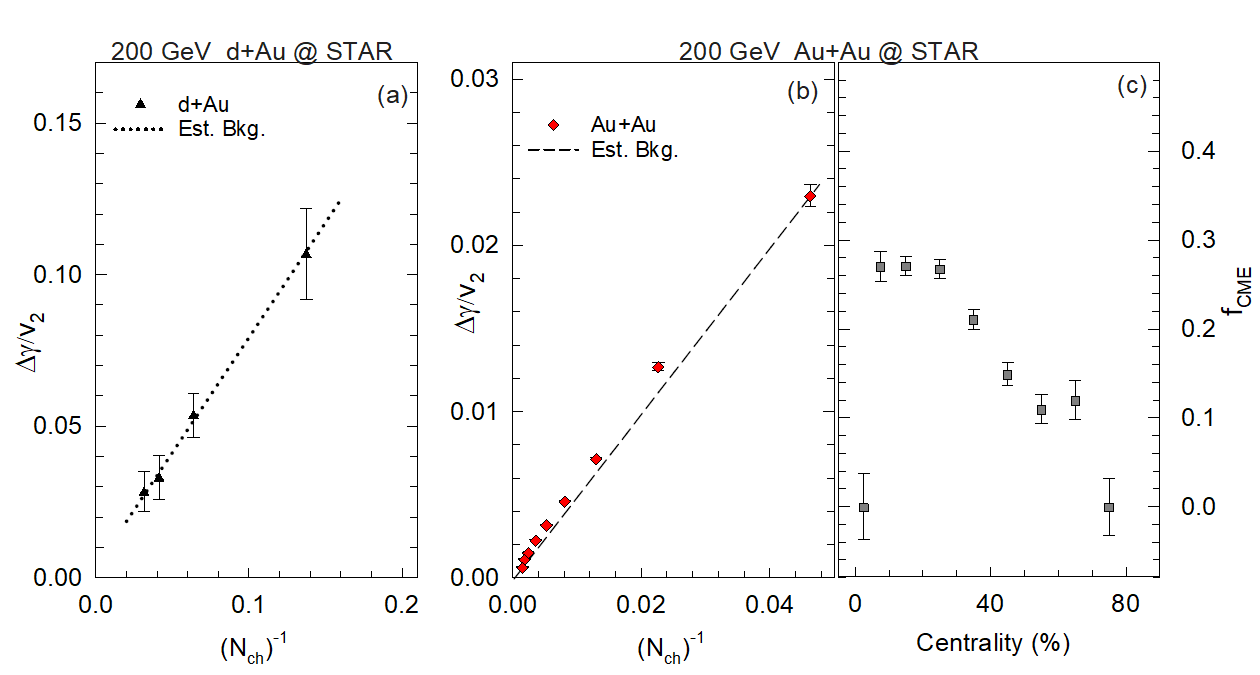}
\vskip -0.10in
\caption{  $\Dg/v_2$ vs. $1/{\rm N_{ch}}$ [(a) and (b)] and $f_{\rm CME}$ vs. centrality (c) for \dA\ and \AuAu\ collisions at $\sqrt s_{\mathrm{NN}}=200$~GeV. The dotted and dashed lines indicate an estimate of the background contributions. 
The data are taken from Refs.~\cite{STAR:2008ftz,STAR:2013ksd,STAR:2019xzd}.
} 
\label{fig3} 
\end{figure*} 
%


The scaling results for collisions at $\sqrt s_{\mathrm{NN}}=200$~GeV are shown in Fig.~\ref{fig3}. The $1/{\rm N_{ch}}$ scaling apparent for \dA\ collisions (Fig.~\ref{fig3} (a)) confirms the expectation that the CME is negligible in these collisions. It also confirms that the combined sources of background (LCC, resonances, back-to-back jets, ...), which could be substantial in \dA\  collisions, show $1/{\rm N_{ch}}$ scaling. In contrast to \dA, the results for \AuAu\ (Fig.~\ref{fig3}(b)) show visible indications of a violation ($c > 0$) to the $1/{\rm N_{ch}}$ scaling observed for background-driven charge separation in $p(d)$+A collisions.  Similar violations were observed for \RuRu\ and \ZrZr\ \cite{Lacey:2022plw}. The scaling violation is similar to that observed for signal + background in Figs.~\ref{fig1} (a) and (c), suggesting an unambiguous non-negligible CME contribution to the measured $\Dg/v_2$ in \AuAu, \RuRu, and \ZrZr\ collisions. The CME contribution can be quantified via fits with the function 
${\Delta\gamma}/{v_2}  = a + {b}/{(\rm N_{ch})^{1- c}},\label{eq:5}$ to evaluate $c$. Alternatively, estimates of the background for all three systems can be estimated by leveraging the $\Dg/v_2$ measurements for peripheral and central collisions with $1/{\rm N_{ch}}$ scaling \cite{Lacey:2022plw}. Here, it is important to recall that the simulated results from the  AVFD and HIJING models, as well as the measurements presented in Figs.~\ref{fig2} and \ref{fig3}(a), provide strong constraints that the combined sources of background, scale as $1/{\rm N_{ch}}$ over the full centrality span. The background estimates were used to extract $f_{\rm CME}$ values for \AuAu, (Fig.~\ref{fig3} (c)) \RuRu\ and \ZrZr\ collisions respectively. They indicate non-negligible $f_{\rm CME}$ values that vary with centrality. In mid-central collisions, $f_{\rm CME} \sim 27\%$ for \AuAu\ collisions, which is roughly a factor of two larger than the values for \RuRu\ and \ZrZr. Within the uncertainties, no significant difference between the  values for \RuRu\ and \ZrZr\ was observed, suggesting that $\Dg/v_2$ is sensitive to CME-driven charge separation in \RuRu\ and \ZrZr\ collisions but may be insensitive to the signal difference  between them \cite{Lacey:2022plw}. 
 
In summary, we demonstrate that the scaling properties of the $R_{\Psi_2}(\Delta S)$ and the $\Delta\gamma$ correlators can be used to characterize the CME in colliding systems at RHIC and the LHC. The $\Dg/v_2$ measurements for \pA\ and \dA\ collisions at $\snn = 200$~GeV and \pPb\ ($\snn = 5.02$~TeV) and \PbPb\  collisions at $\snn = 5.02$ and $2.76$ TeV, scales as $1/{\rm N_{ch}}$ consistent with background-driven charge separation. However,  the results for \AuAu, \RuRu\ and \ZrZr\ collisions ($\snn = 200$~GeV) show scaling violations, which indicate a CME-driven contribution in the presence of significant background. In mid-central collisions, $f_{\rm CME} \sim 27\%$ for \AuAu\ collisions and is approximately a factor of two smaller in \RuRu\ and \ZrZr\ collisions with similar magnitudes for the two isobars.



%
%
%

\end{document}